\documentclass[aps,prl,twocolumn,showpacs,amsmath,superscriptaddress,floatfix]{revtex4}

\usepackage{amsfonts}
\usepackage{amsmath}
\usepackage{amssymb}
\usepackage{graphicx}
\usepackage{color}
\usepackage{verbatim}
\usepackage{lipsum}
\usepackage{dsfont}
\usepackage{hyperref}
\usepackage{upgreek}

\begin{document}
\title{Trapping and sympathetic cooling of single thorium ions for spectroscopy}

\author{Karin Groot-Berning}
\email{karin.groot-berning@uni-mainz.de}
\affiliation{QUANTUM, Institut f\"ur Physik, Universit\"at Mainz, Staudingerweg 7, 55128 Mainz, Germany}

\author{Felix Stopp}
\affiliation{QUANTUM, Institut f\"ur Physik, Universit\"at Mainz, Staudingerweg 7, 55128 Mainz, Germany}

\author{Georg Jacob}
\affiliation{QUANTUM, Institut f\"ur Physik, Universit\"at Mainz, Staudingerweg 7, 55128 Mainz, Germany}

\author{Dmitry Budker}
\affiliation{QUANTUM, Institut f\"ur Physik, Universit\"at Mainz, Staudingerweg 7, 55128 Mainz, Germany}
\affiliation{Helmholtz Institut, Universit\"at Mainz, 55128 Mainz, Germany}
\affiliation{Department of Physics, University of California, Berkeley, California 94720-7300, USA}
\affiliation{PRISMA Cluster of Excellence, Johannes Gutenberg Universit\"at Mainz, 55128 Mainz, Germany}

\author{Raphael Haas}
\affiliation{Helmholtz Institut, Universit\"at Mainz, 55128 Mainz, Germany}
\affiliation{Institut f\"ur Kernchemie, Universit\"at Mainz, 55128 Mainz, Germany}

\author{Dennis Renisch}
\affiliation{Helmholtz Institut, Universit\"at Mainz, 55128 Mainz, Germany}
\affiliation{Institut f\"ur Kernchemie, Universit\"at Mainz, 55128 Mainz, Germany}

\author{J\"org Runke}
\affiliation{Institut f\"ur Kernchemie, Universit\"at Mainz, 55128 Mainz, Germany}
\affiliation{GSI Helmholtzzentrum f\"ur Schwerionenforschung GmbH, 64291 Darmstadt, Germany}

\author{Petra Th\"orle-Pospiech}
\affiliation{Helmholtz Institut, Universit\"at Mainz, 55128 Mainz, Germany}
\affiliation{Institut f\"ur Kernchemie, Universit\"at Mainz, 55128 Mainz, Germany}

\author{Christoph D\"ullmann}
\affiliation{Helmholtz Institut, Universit\"at Mainz, 55128 Mainz, Germany}
\affiliation{PRISMA Cluster of Excellence, Johannes Gutenberg Universit\"at Mainz, 55128 Mainz, Germany}
\affiliation{Institut f\"ur Kernchemie, Universit\"at Mainz, 55128 Mainz, Germany}
\affiliation{GSI Helmholtzzentrum f\"ur Schwerionenforschung GmbH, 64291 Darmstadt, Germany}

\author{Ferdinand Schmidt-Kaler}
\affiliation{QUANTUM, Institut f\"ur Physik, Universit\"at Mainz, Staudingerweg 7, 55128 Mainz, Germany}
\affiliation{Helmholtz Institut, Universit\"at Mainz, 55128 Mainz, Germany}
\affiliation{PRISMA Cluster of Excellence, Johannes Gutenberg Universit\"at Mainz, 55128 Mainz, Germany}

\begin{abstract}
Precision optical spectroscopy of exotic ions reveals accurate information about nuclear properties such as charge radii and magnetic and quadrupole moments. Thorium ions exhibit unique nuclear properties with high relevance for testing symmetries of nature. We report loading and trapping of single $^{232}$Th$^+$ ions in a linear Paul trap, embedded into and sympathetically cooled by small crystals of trapped $^{40}$Ca$^+$ ions. Trapped Th ions are identified in a non-destructive manner from the voids in the laser-induced Ca fluorescence pattern emitted by the crystal, and alternatively, by means of a time-of-flight signal when extracting ions from the Paul trap and steering them into an external detector. We have loaded and handled a total of 231 individual Th ions. We reach a time-of-flight detection efficiency of $\gtrsim 95\, \%$, consistent with the quantum efficiency of the detector. The sympathetic cooling technique is expected to be applicable for other isotopes and various charge states of Th e.g., for future studies of $^{229m}$Th. 
\end{abstract}

\maketitle
The low-lying isomeric excitation in $^{229}$Th is a unique case where a nuclear transition may be accessible by direct laser excitation. Recently, $^{229m}$Th deexcitation to the ground state was directly observed via detection of conversion electrons emitted in the decay of neutral $^{229m}$Th \cite{wense_direct_2016}, and the half-time in neutral $^{229m}$Th was determined to be 7(1)\,$\mu$s \cite{seiferle_lifetime_2017}. Numerous applications of this low-lying isomer have been proposed, including a nuclear laser \cite{tkalya_proposal_2011} and nuclear quantum optics \cite{burvenich_nuclear_2006}, as well as its use as a reference for an optical clock of unprecedented accuracy \cite{raeder_resonance_2011,peik_nuclear_2003,campbell_single-ion_2012,wense_towards_2018}. Such a clock is of interest as a tool to search for dark matter \cite{derevianko_hunting_2014} and gravitational waves \cite{Kolkowitz2016}, as well as in relativistic geodesy \cite{Bondarescu2015}. It also promises ultra-high sensitivity to time variations of fundamental constants \cite{flambaum_enhanced_2006, thielking_laser_2018} making the $^{229/229m}$Th system to be of great interest for both fundamental science and applications. 

The first characterization of the $^{229m}$Th nucleus via hyperfine spectroscopy \cite{thielking_laser_2018} revealed discrepancies between the experimentally determined value of the magnetic moment of the isomeric state and theoretical estimates based on the Nilsson deformed-shell model, which were attributed to neglecting  effects like collective quadrupole-octupole coupling. More precise spectroscopic measurements are needed to resolve the uncertainties about the sensitivity of the $^{229}$Th system to the variation of fundamental constants \cite{berengut_proposed_2009,thielking_laser_2018}.

In the present work, we aim to develop and exploit single-ion trapping and spectroscopy of $^{232}$Th$^+$. Trapped single ions or linear crystals with only a few ions allow for cooling to temperatures in the $\mu$K regime, which, in combination with near-to-perfect micro motion compensation \cite{KELLER2018}, leads to much reduced systematic frequency shifts. Eventually ions and ion crystals are cooled to the motional ground state \cite{LECH2016} and offer options for quantum logic spectroscopy \cite{SCHM2005}, a technique which has been used for frequency measurements exceeding relative accuracy of 10$^{-17}$ \cite{CLOCKS2015}. However, implementation of such advanced techniques to exotic ions meets additional challenges as one needs to trap, handle and also identify ions at a single-particle level.

Previous work on ion trapping of Th includes production of laser-cooled crystals of $^{232}$Th$^{3+}$ \cite{campbell_multiply_2009} and $^{229}$Th$^{3+}$ \cite{campbell_wigner_2011}  in a linear Paul trap, which enabled accurate spectroscopy of electronic states. Clouds of 10$^6$ buffer-gas cooled $^{232}$Th$^+$ ions have also been trapped in a macroscopic linear Paul trap and used for precision spectroscopy of electronic states \cite{herrera-sancho_energy_2013}.

\begin{figure*}[h!tp]
\begin{center}
\includegraphics[width=0.8\textwidth]{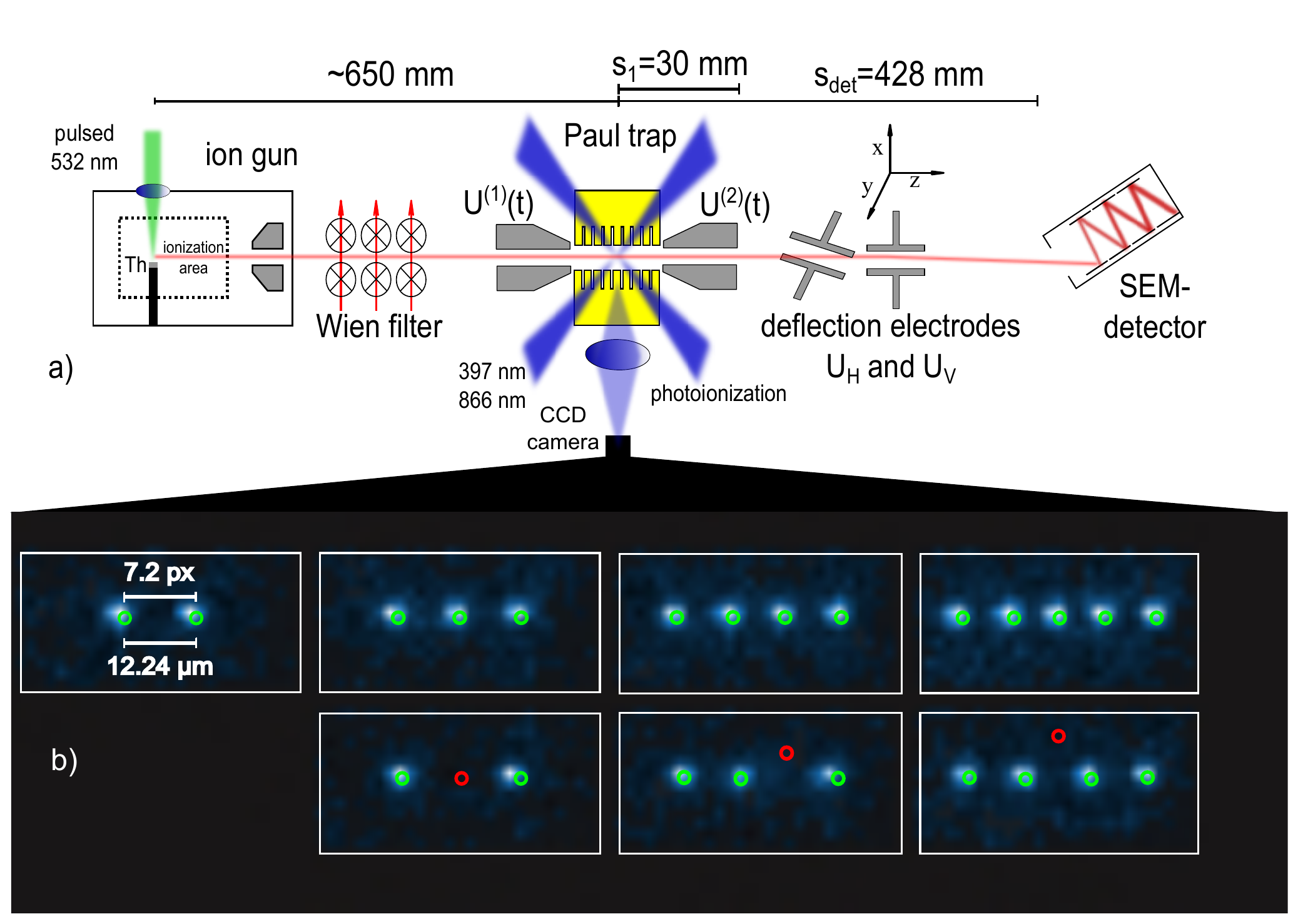}
\caption{(a) Sketch of the experimental setup, (b) Observation of the laser-induced Ca-ion fluorescence for crystals with and without embedded Th and different number of Ca ions. Upper row: Calibration of the distance between two ions: at a measured axial trap frequency of 310\,kHz, it is 12.24\,$\upmu$m, corresponding to 7.2 pixel (px). The exposure time is 40\,ms. Ca ion crystals arrange in linear configuration. Lower row: Observation of mixed-species crystals. Th ion trapped at the location of dark void. Calculated ion positions for Ca (green) and Th (red) are indicated with circles, assuming the trap frequencies of $\omega_{x,y,z}=2\pi\cdot$ (1640, 1758, 310)\,kHz.}
\label{fig:Setup}
\end{center}
\end{figure*}

Here we provide a full set of procedures and specialized instrumentation for trapping and sympathetically cooling single Th ions and demonstrate their identification either non-destructively from the ion-crystal fluorescence, when Th ions are trapped in a linear $^{40}$Ca$^+$ ion crystal, or alternatively, by means of a time-of-flight signal from extracting single Th ions out of the trap into a secondary-electron-multiplier (SEM)  detector downstream. The experimental setup comprises three parts, a source of Th ions, an ion trap for sympathetic cooling and in-situ detection, and a time-of-flight (TOF) detector [Fig.\,\ref{fig:Setup}(a)].

The source of Th ions is based on a commercial ion gun (Specs IQE 12/38) where we inserted a ceramic rod to hold a sample of Th metal (diameter 3\,mm, thickness 89\,$\upmu$m) of 99.85\,$\%$ certified purity (Reactor Experiments, Inc., San Carlos, CA, USA) with maximum of Fe (0.13\,$\%$), Si (0.015\,$\%$), Sn (0.001\,$\%$), Ni (0.001\,$\%$), and further concentrations $\le$0.001\,$\%$ of Cu, Cr,  Mn, Mg, Ca, which is ablated a single pulse from a frequency-doubled Nd:YAG laser. 

The ablated Th atoms are ionized inside the volume enclosed by the repeller electrode structure (see Fig.~1(a), dotted), accelerated by an  extraction electrode (grey) and separated by a Wien filter. The beam is injected into a linear Paul trap. The ions are steered through a central hole (200\,$\upmu$m diameter) in the endcap (28.55\,mm length) of the trap. The ion trap features two endcaps of identical shape at a distance of 2.9\,mm, and RF- and segmented DC-electrodes in a X-blade design \cite{JAC2016}. The transversal size of the trap is 960\,$\upmu$m, given by the distance between opposite electrodes. We operate the RF electrodes with a radio frequency of $\Omega$~=~2$\pi\cdot $ 23.062\,MHz with a peak-to-peak amplitude of U$_{\text{pp}}$=572\,V, generating a harmonic radial pseudopotential of $\omega_{x,y}$ = 2$\pi\cdot $(1.640, 1.758)\,MHz. The axial confinement in the trap is achieved with voltages U$^{(1),(2)}(t)$ which are applied to both endcaps, in combination with control voltages on the 11 DC-blade segments from a home-built multichannel arbitrary waveform generator. Addressing the endcaps and the DC segments with time-dependent voltage ramps allows for a versatile control of the capture, confinement, and ion extraction from the trap. 

Th ions are injected from the ion gun source, while Ca is loaded by photoionization of an atomic beam emerging from a resistively heated oven (not shown in Fig.~\ref{fig:Setup}). For continuous laser cooling of Ca$^+$ we use light 20\,MHz red detuned from the S$_{1/2}$ to P$_{1/2}$ transition near 397\,nm, while resonant laser radiation tuned around 866\,nm and 854\,nm is used for repumping from the metastable D-levels. Laser-induced fluorescence at 397\,nm wavelength is imaged via a f/1.76 optics with magnification of 14.1 onto the chip of an intensified camera (EMCCD). We determine the magnification of the optical system from the distance between the Ca$^+$ ions in a two-ion crystal which is deduced from the measured ions' trap frequency, see Fig.~\ref{fig:Setup}(b). This results in a calculated equilibrium distance of 12.24\,$\upmu$m \cite{JAMES1998}. As the spatial resolution in imaging ions within crystals is below 0.5\,px, crystal structures are well resolved, and ion positions can be determined with a 0.6\,$\upmu$m precision. Image acquisition, counting the SEM, and the application of voltage ramps is executed with an experiment-control system with ns-time resolution. Once the Th ions are trapped, sympathetically cooled and identified, they may be extracted from the trap via the pierced endcap on the exit side of the trap and steered into the SEM detector. It features a pulse width of 6.2\,ns (FWHM), and an independently measured detection efficiency of 0.96(2) for ions with a few keV energies.  

\begin{figure}[t]
\centering
\includegraphics[width=0.89\columnwidth]{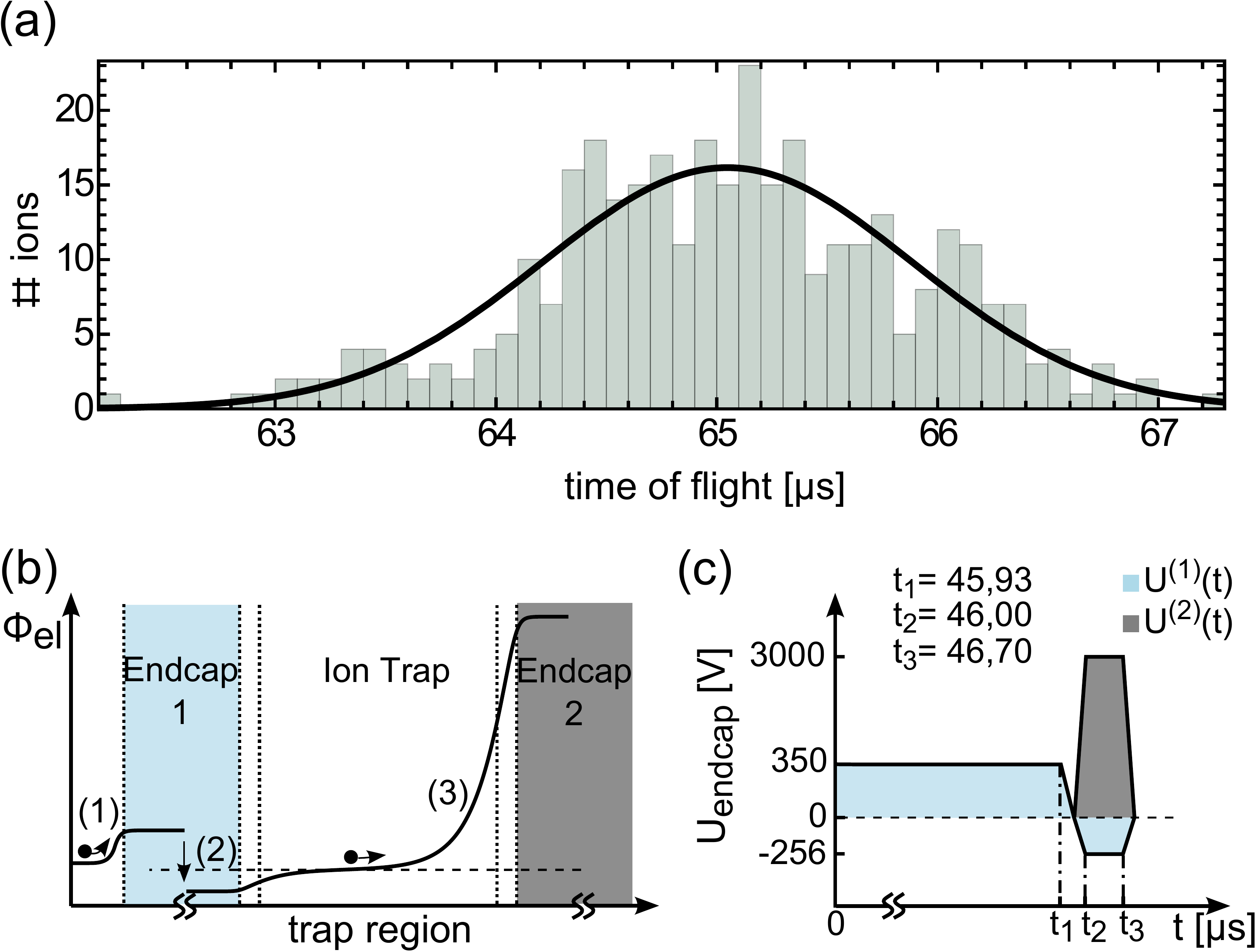}
\setlength{\belowcaptionskip}{5ex}
\caption{(a) Measured TOF distribution of laser-ablated and ionized Th ions from the ion gun. (b) Sketch of the time-dependent electric potential for ion loading. (1) Incoming ions are decelerated by a repulsive potential $U^{(1)}$ = +350\,V. When they are inside the first endcap, its potential is switched to $U^{(1)}$= -256\,V and the ions get decelerated a second time since they do not see a potential change within the endcap. (3) The second endcap is switched to $U^{(2)}$= +3\,kV simultaneously thus the ions are repelled back into the trap volume.  (c) Timing sequence for deceleration of incoming Th ions for trap loading.}
\label{fig:loading}
\end{figure}

The experimental sequence for trapping, cooling and identifying Th ions is as follows: Th ions are generated with a short laser pulse (pulse width 5(2)\,ns, 1.7\,mJ) at 532\,nm and extracted from the ion gun. Note that the laser pulse just ablates neutral Th, and we rely on the ionization by electron impact in the ion gun. This is different from the recent work on mass spectroscopy with different charge states of Th, where ion clouds were generated by plasma ionization \cite{Borisyuk2017}. When we extract ions at an energy of 600\,eV we obtain a velocity distribution from a TOF signal, from steering the beam through the endcap holes of the ion trap but keeping the endcaps at ground potential. A spread of a TOF distribution of about 2\,$\upmu$s (FWHM) is determined, see Fig.\,\ref{fig:loading}(a), from which we estimate a spatial width of the incoming ions of 20\,mm at the position of the entrance ion trap endcap. 

The Th ions are decelerated and captured using time-dependent voltages on the endcaps. For this, incoming ions are exposed to a repulsive potential from endcap 1 at $U^{(1)}$ = +350\,V, chosen such that the ion energy is sufficient to enter the hole of endcap 1. Now, the endcap is switched to attractive $U^{(1)}$= -256\,V. 
While ions inside the hole are shielded from the electric field and do not see the potential change, the Th ions get decelerated a second time, when exiting endcap 1.  Simultaneously, the voltage at the opposite endcap $U^{(2)}$= +3.0\,kV is set to a repulsive potential repelling the ions back into the trap volume. Finally, both endcaps are switched to ground potential and only the DC-segments are used for forming a tightly confining axial potential with a trap frequency of  $\omega / (2\pi)$=309.9\,kHz. The sequence of switching the endcap potentials was initially optimized using numeric simulations of ion trajectories, see Fig.\,\ref{fig:loading}(b), taking into account the full time-dynamics of the Paul potential and a realistic geometric model of the ion trap\,\cite{SINGER2010}.
From this simulation and with further optimization on the experimental setup, we find the voltages and timings given in Fig.\,\ref{fig:loading}(c).

In the experiments, we trap up to $n$=5 Ca ions. Under typical operating conditions, Ca ions crystallize in a linear shape with inter-ion distances of 8.0 to 12.2\,$\upmu$m. Equilibrium positions are determined from fits to the ion fluorescence, and we find good agreement with calculated predictions\,\cite{JAMES1998}. For crystals with embedded Th, voids in the fluorescence between bright Ca ions indicate the location of Th ``impurities.''. In case of $n$=2 Ca ions and one Th, the trap anisotropy is sufficiently large, characterized by an $\alpha=\omega_{\text{ax}}/\omega_{\text{rad}}$= 0.19, to keep the mixed crystal in a linear structure. However, for $n$=3 and 4, the linear symmetry of the crystal is broken, and we find the Th ion generating a topological defect\,\cite{PART2013} indicating the transition to a zigzag structural phase\,\cite{KAUF2012}. This is confirmed by crystal-structure calculations which take into account the measured trap frequencies and reproduce the fitted ion positions from the imaging with a relative accuracy of better than 5\,$\%$. Note that ions in a different charge state\,\cite{FELD2014} or ions with different charge-over-mass ratio as compared to Th$^+$ would result in different ion positions. Assuming an uncertainty of 10\,kHz in determining axial and radial trap frequencies, observation of a structural phase transition breaking the linear symmetry of an $n$=2 crystal would indicate a mass of $m \geq$ 285(5), while an $n$=3 crystal would assume a linear configuration for $m \leq$ 230(5). Observation of structural phase transitions by adjusting the trap anisotropy is, in fact, a versatile tool for in-situ identification of dark trapped ions.
\begin{figure}[t]
\centering
\includegraphics[width=0.95\columnwidth]{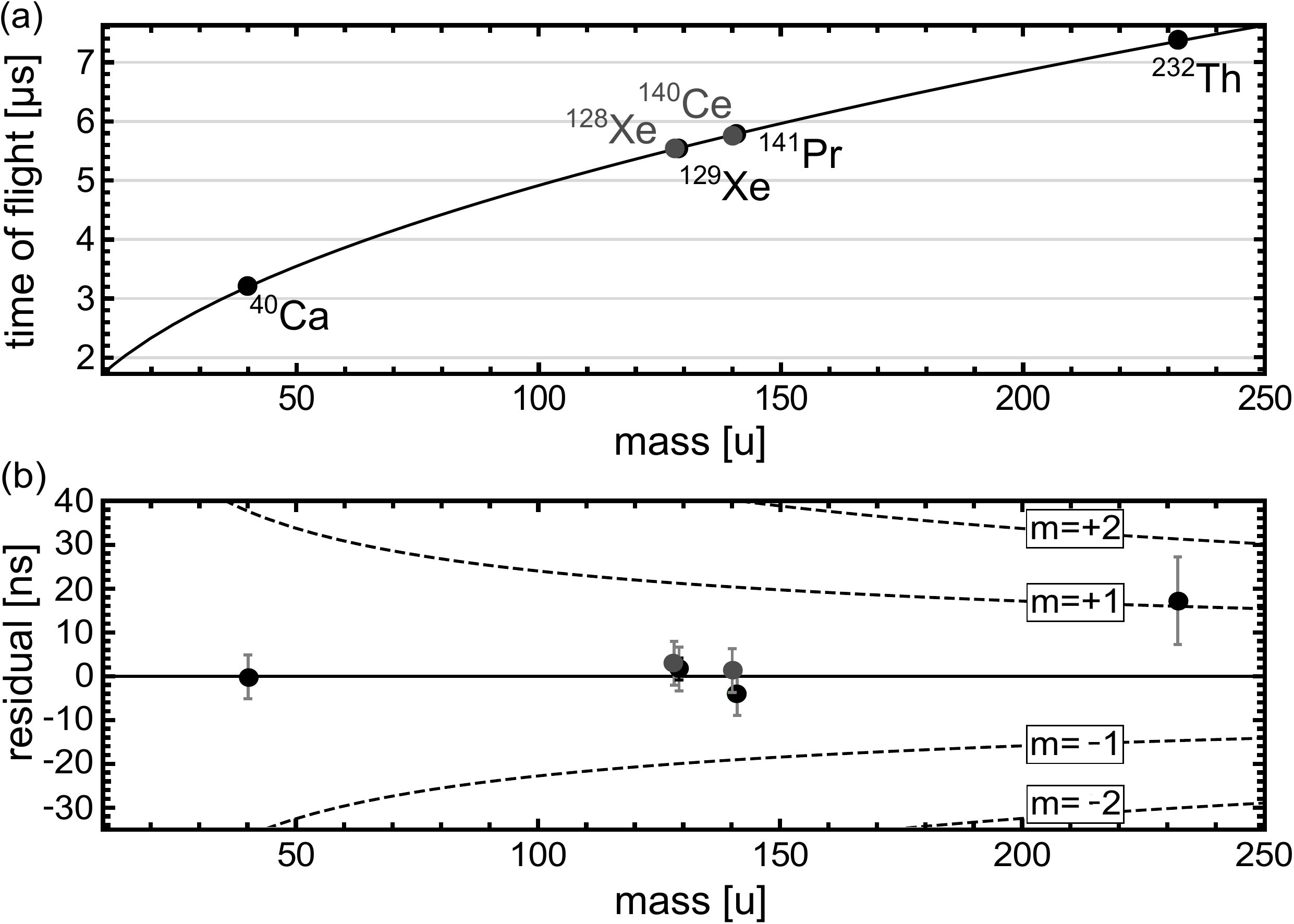}
\setlength{\belowcaptionskip}{5ex}
\caption{(a) TOF identification for ions extracted from the linear ion trap. Note that data points for isotopes of Xe and Pr/Ce in gray and black, are partially overlapping.  (b) Residuals of the TOF identification, including systematic errors due to the path into the detector, see text and Fig.\,\ref{fig:errors}. Total errors are 10\,ns for Th ions and and 5\,ns for all other ion species.}
\label{fig:TOF}
\end{figure}

Such non-destructive identification of trapped Th is confirmed by TOF experiments. For this, we apply an attractive potential $U^{(2)}_\text{i}$=-2.9\,kV to the exit endcap for 300\,ns. After all Th and Ca ions are extracted into the endcap hole, the potential is reverted to $U^{(2)}_\text{f}$= +3.0\,kV such that ions exiting the hole are further accelerated, steered by the deflection electrodes with voltages $U_H$ and $U_V$, respectively, and hitting the SEM detector, see Fig. 1(a). 
Endcap 1 is held at a constant $U^{(1)}$=0\,V. We model the measured TOF-time $t_{\text{meas}}(U^{(2)}_\text{i}, U^{(2)}_\text{f})$ of ions with  
\begin{equation}
t_{\text{meas}}=\sqrt{\frac{m s_1^2}{2 e k_1 U^{(2)}_\text{i}}} + \sqrt{\frac{m (s_\text{det} - s_1)^2}{2 e (k_1 U^{(2)}_\text{i} + k_2 U^{(2)}_\text{f})}} + t_{\text{del}}, 
\label{equ:TOF}
\end{equation}
where $m$ is the mass of the ion, $U^{(2)}_\text{i,f}$ voltages applied to endcap 2, $s_1$=\,30\,mm and $s_{\text{det}}$=\,428\,mm are the distances from the ion-trap center to the endcap, and from the ion-trap center to the detector, respectively. 

\begin{figure}[t]
\centering
\includegraphics[width=0.80\columnwidth]{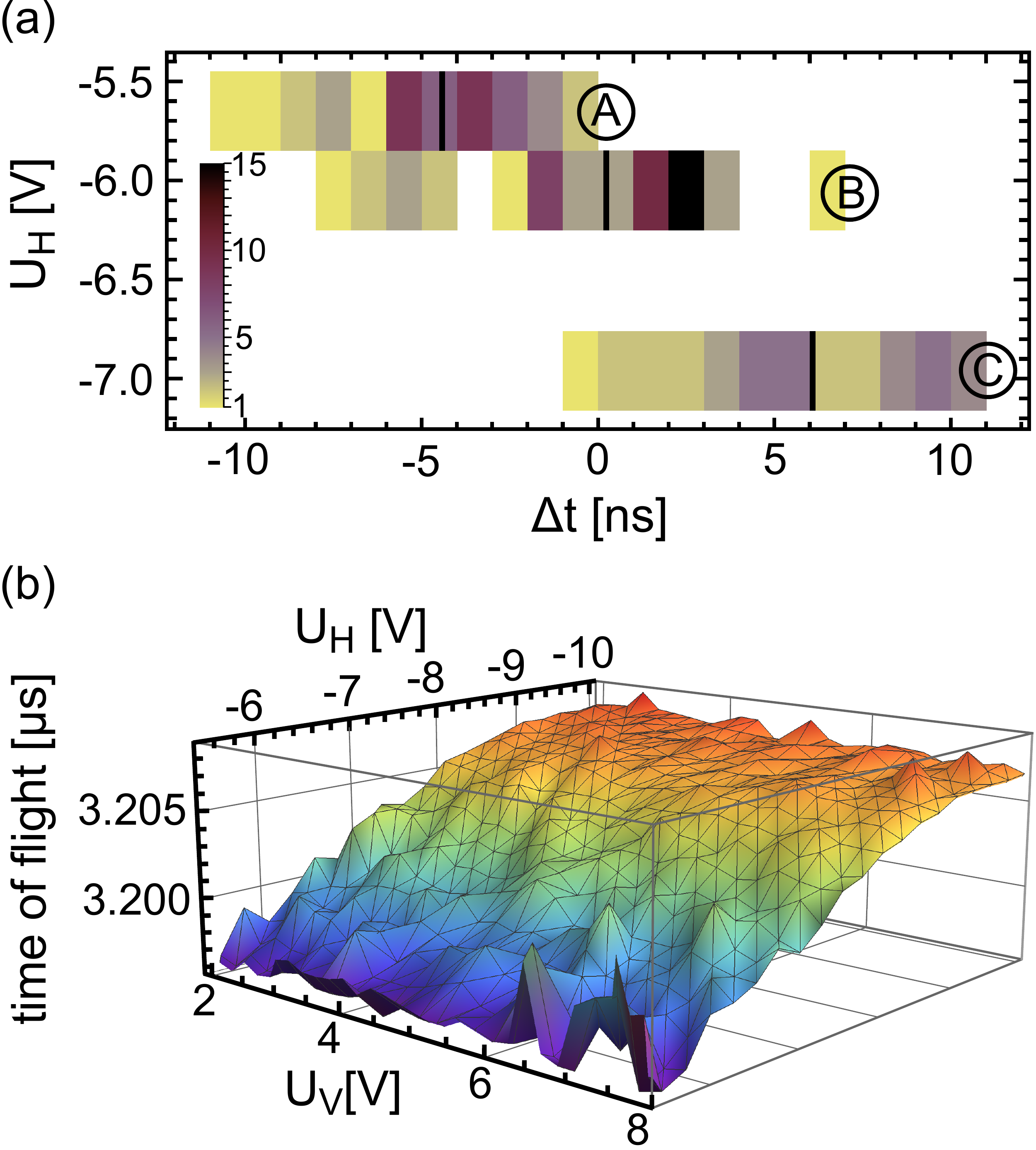}
\setlength{\belowcaptionskip}{5ex}
\caption{(a) Determination of the systematic TOF error. Th ions are steered, by a variation of the deflection voltage, onto different spots of the detector, giving rise to a variation of the TOF signal. Histograms for Th steered into a horizontal detector corner (A), the center (B),  and onto the position corresponding to the center of the detector as determined using Ca ions (C).  (b) Measured TOF variation as Ca ions are steered over the entire detector area in the vertical and horizontal directions.}
\label{fig:errors}
\end{figure}

The TOF data from extracting ions of different mass, $^{40}$Ca$^+$, $^{128}$Xe$^+$, $^{129}$Xe$^+$, $^{140}$Ce$^+$, and $^{141}$Pr$^+$ with $U_{\text{i,f}}$ of -2.9\,kV, +3\,kV, respectively, are shown in Fig.\,\ref{fig:TOF}(a) together with the model curve, and obtaining the fit parameters $k_{1,2}=\{0.709(3),\,0.912(2)\}$ and $t_{\text{del}}$=246(6)\,ns. In case of Ca we extract a single ion and in case of the other isotopes we extract exact one of those plus another single Ca ion.
The measured TOF data for $^{232}$Th$^+$ agrees with the fit. The errors shown in Fig.\,\ref{fig:TOF}(b) include statistical uncertainties which are larger for Th due to a smaller number of events, as compared to the other species, and a larger width of the TOF distribution. This is because we here extract ion crystals with up to 6 ions, including mixed-species zigzag formations.  Another  error comes from the off-axis SEM detector mounted in a tilted way, to use it as an on-axis detector. The dominant systematic error stems from the variation of the apparent TOF when ions hit different positions of the $\sim$5.5\,mm diameter entrance hole of the SEM. We investigated this by horizontally deflecting the Th impact point, and observing an overall shift in the TOF signal equivalent to about 1.38 mass units, see Fig.\,\ref{fig:errors}(a). This finding is further confirmed by mapping of the TOF over the detector entrance area using single Ca ions under identical conditions as we used for the Th. We find TOF variations of about 10\,ns, with the identical systematic shift to longer TOF for larger deflection voltages U$_H$. The limited accuracy for the single-beam alignment on the detector center dominates the systematic-error budget.

Even when aligning the Th beam into the center of the detector, we conjecture a difference in fine adjustment and beam propagation inside the detector to explain the TOF offset from the model function, and estimate a systematic error. Note that such error could be mitigated by detectors with higher temporal resolution and smaller aperture. Nevertheless, in our experiment single Th ions are unambiguously identified at the single-particle level and are counted with better than 95\,$\%$ efficiency.\\

In conclusion, we have demonstrated capturing and loading single Th ions into a Paul trap and sympathetically cooling them in crystals of cold Ca ions. The combination of in-situ detection from the ion-crystal fluorescence and from destructive TOF detection has been clearly shown. This establishes new options for single-ion spectroscopy with exotic ions, eventually quantum logic spectroscopy using the Ca as a read-out of quantum states and transitions of Th ions which escape direct observation via detection of fluorescence or which cannot be interrogated using the electron-shelving method. In the future, we plan to employ the advanced quantum-computing\,\cite{KAUF2017} and quantum-sensing\,\cite{RUST2017}  techniques with entangled Ca ion crystals for precise investigations on the unique system of $^{229}$Th$^+$ and $^{229\text{m}}$Th$^+$.

This work was financially supported by the Helmholtz Excellence Network ExNet-0020, Precision Physics, Fundamental Interactions and Structure of Matter (PRISMA+) from the Helmholtz Initiative and Networking Fund, and we acknowledge financial support by the DFG DIP program (FO 703/2-1) and by the VW Stiftung. 

\bibliography{Ions,Thorium}
\bibliographystyle{apsrev4-1}

\end{document}